\documentclass[conference]{IEEEtran}
\IEEEoverridecommandlockouts
\usepackage{graphicx}
\usepackage[table,xcdraw,dvipsnames]{xcolor}
\usepackage{tabularx}
\usepackage{ulem}
\usepackage{tabularray}
\usepackage{multirow}
\usepackage[bottom]{footmisc}
\usepackage{cite}
\usepackage{amsmath,amssymb,amsfonts}
\usepackage{cleveref}
\usepackage{algorithmic}
\usepackage{textcomp}
\usepackage{flushend}
\usepackage{wrapfig}

\def\BibTeX{{\rm B\kern-.05em{\sc i\kern-.025em b}\kern-.08em
    T\kern-.1667em\lower.7ex\hbox{E}\kern-.125emX}}
\newcommand{\url}[1]{\small{\textit{{#1}}}}
\newcommand{\todo}[1]{}

\newcommand{\hide}[1]{#1}
\usepackage{tikz}

\makeatletter 
\newcommand{\linebreakand}{%
  \end{@IEEEauthorhalign}
  \hfill\mbox{}\par
  \mbox{}\hfill\begin{@IEEEauthorhalign}
}
\makeatother 

\begin{document}

\title{Advocate - Trustworthy Evidence in Cloud Systems\thanks{
Acknowledgements: Thanks also to Luka Banse, Mathis Kähler, Kevin Pham and Felix Graf for their work in building and testing Advocate.
Funded by the European Union (TEADAL, 101070186). Views and opinions expressed are, however, those of the author(s) only and do not necessarily reflect those of the European Union. Neither the European Union nor the granting authority can be held responsible for them.
}}%

\hide{
\author{\IEEEauthorblockN{Sebastian Werner, Sepideh Masoudi, Fernando Castillo, Fabian Piper, Jonathan Heiss}
\IEEEauthorblockA{\textit{Information Systems Engineering}, 
\textit{Technische Universität Berlin}, Germany \\
\{sw,smi,fc,fpi,jh\}@ise.tu-berlin.de}
}
}

\maketitle
\thispagestyle{plain}
\pagestyle{plain}

\begin{abstract}
The rapid evolution of cloud-native applications, characterized by dynamic, interconnected services, presents significant challenges for maintaining trustworthy and auditable systems, especially in sensitive contexts, such as finance or healthcare.
Traditional methods of verification and certification are often inadequate due to the fast-past and dynamic development practices common in cloud computing.
This paper introduces Advocate, a novel agent-based system designed to generate attested evidence of cloud-native application operations. 
By integrating with existing infrastructure tools, such as Kubernetes and observability services, Advocate captures, authenticates, and stores evidence trails in a tamper-resistant manner. 
This approach not only supports the auditing process but also allows for privacy-preserving evidence aggregation.
Advocate’s extensible architecture facilitates its deployment in diverse environments, enabling the verification and adherence to policies and enhances trust in cloud services.
\end{abstract}

\section{Introduction}

Cloud-native applications are characterized as highly dynamic, quickly iterating applications that often consist of several interdependent and interconnected services.
These services are often a mixture of managed and self-developed components that together deliver highly scalable, available, and elastic applications using public cloud providers.
Consumers of these applications have difficulty verifying or understanding how such an application delivers responses. 
It can be unclear which version of an application was used, in which environment (hardware) this application ran on, and which other cloud services were used. 
Such information may become critical in sensitive contexts such as finance, insurance, or healthcare.
Application consumers may require evidence about the application and its execution to be presented to third parties, for example, for billing purposes or to demonstrate regulatory compliance.
The requirement to provide trustworthy and auditable software clashes with current capabilities of modern cloud-native applications. 
Classical approaches, such as formal verification or software process certifications are too time-consuming and may not extend to the cloud providers' infrastructure.

To address this dichotomy of dynamic fast-paced application operation and auditable trustworthy applications we propose Advocate -- an \textbf{A}gent for \textbf{D}ata \textbf{V}erification, \textbf{O}n-\textbf{C}haining and \textbf{A}nchoring in \textbf{T}rustless Environments.
Advocate is an agent-based system designed for integration into cloud-based environments. 
It introduces a novel approach that acknowledges the inherently imperfect and dynamic nature of cloud-native applications while providing a best-effort method for generating verifiable evidence to audit and analyze an application's behavior, 
following the TrustOps\cite{trustOps} framework. 


\label{sec:intro}
\section{Design Objectives}

\noindent Below, we outline Advocate's objectives and design choices.

\textbf{R1: Cloud-native}: 
Cloud-native environments consist of many different, managed, and self-configured services.
This requires comprehensive monitoring and observability capabilities including each part of an application. 
In Advocate, we address this through an agent-based, easily integrable, and extensible design, technically realized through tight integration into Kubernetes at deployment time. 

\textbf{R2: Holistic Integration}:
Relevant operations originate from both, the running applications, e.g., incoming and outgoing requests, and their execution infrastructure, e.g., deployment of virtual machines for scalability.
Providing a seamless and comprehensive record of application deployment and execution requires observations on the application and infrastructure levels.
For that, Advocate relies on state-of-the-art observability tools that collectively allow for full-stack observations and generations of evidence.  

\textbf{R3: Independent Verifiability}:
Collected evidence should help application consumers and third-party stakeholders obtain insights into the hosting and execution of applications to help prove certain claims, e.g., resource consumption for billing or carbon footprinting. 
For that, we adopt an SSI-like attestation model where Advocate serves as an \textit{issuer} that attests to credentials (events, settings, ...) of applications that represent credential \textit{holders}. 
This results in attested evidence that can be presented to and validated by third-party \textit{verifiers}. 

\textbf{R4: Presentability and Accessibility}:
Cloud-native applications generate large amounts of potentially sensitive events during their lifetime, hence large amounts of evidence, which requires a suitable storage and access model.
Advocate supports capabilities to realize aggregation and policy-driven transformations on the event data. 
This helps bring the collected evidence into a format that facilitates management and allows for privacy-preserving presentations to third-party \textit{verifiers}.

\textbf{R5: Tamper-Resistance and Availability}:
In multi-party environments like industrial consortiums, applications are often distributed across mutually non- or semi-trusted parties. 
To prevent tampering of evidence in hindsight and improve availability guarantees, Advocate combines distributed ledger technology (DLT) with content-addressable storage (CAS) systems. 
Cryptographic commitments to evidence records are stored and managed by DLT-based smart contracts whereas the plain data is stored in the CAS system as described in~\cite{eberhardt2018off}.

\label{sec:main}
\section{Advocate: Architecture and Implementation}
\begin{figure}
    \centering
    \includegraphics[width=1.05\columnwidth,trim={7 7 7 7},clip]{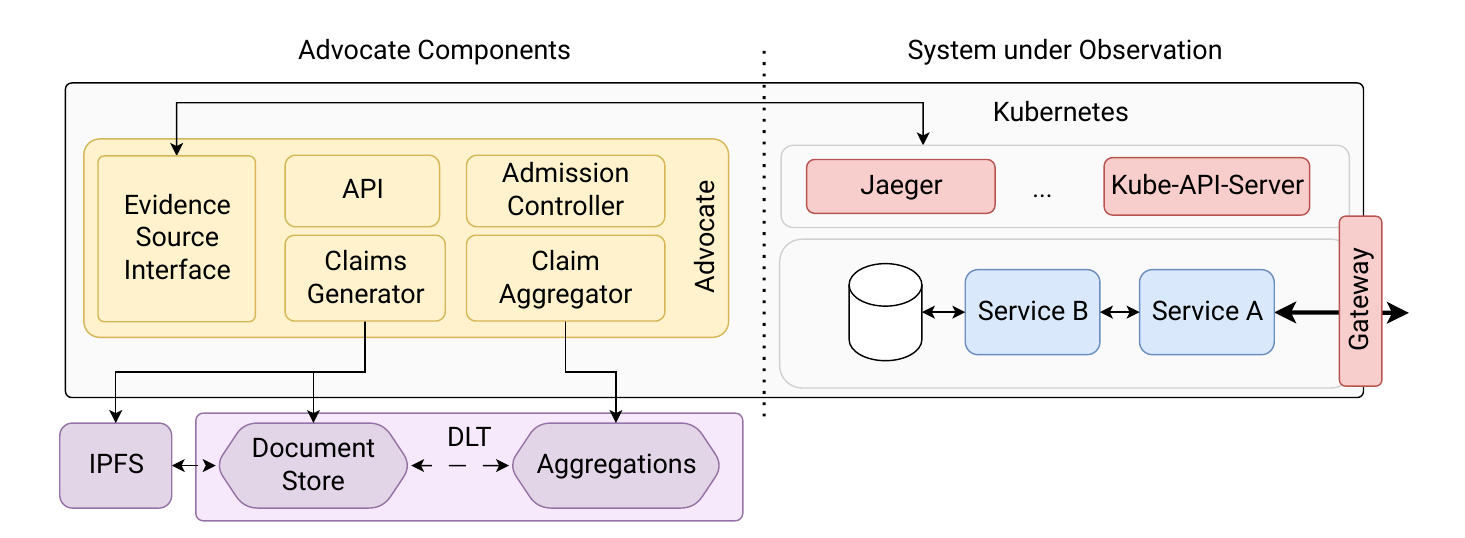}
    \caption{System architecture of Advocate (yellow), Evidence Sources (red), Observed Application (blue), Decentralized Evidence Stores (purple).}
    \label{fig:arch}
\end{figure}

Advocate\footnote{Available online here: \url{https://git.tu-berlin.de/teadal/advocate\textunderscore pie}.} is build as an extensible Python agent that is for now tailored towards Kubernetes environments (\textbf{R1}). 
We assume that these environments are hardened and follow good access practices to reduce the risk of malicious use ideally with verifiable deployments, i.e., through NIX\footnote{\url{https://nixos.org/}}. \Cref{fig:arch} shows the general architecture of Advocate and how it is designed to handle a diverse range of evidence sources and types, e.g., Jaeger (a distributed tracing engine) or an Istio Gateway.
By supporting various correlated evidence sources from different providers and integrating seamlessly with existing infrastructure Advocate ensures that interactions with the applications represent the most comprehensive evidence trail possible (\textbf{R2}).
All evidence sources are either collected event-driven or on pull-based time intervals. 
We provide a simple interface to add evidence sources.
Lastly, to further enrich the collected evidence, Advocate deeply integrates into Kubernetes by using a \textit{AdmissionController} that intercepts and modifies all resources deployed in the cluster. 
This way, Advocate is able to inject additional metadata into each deployed service and collect evidence about how and what services are deployed (\textbf{R1}).

All the collected evidence is turned into verifiable credentials by the \textit{Claims generator} and stored in a decentralized storage system \textit{IPFS}. If configured, this evidence is also directly anchored in an \textit{DocumentStore} smart contract based on OpenAttestation\footnote{\url{https://github.com/Open-Attestation/document-store}} (\textbf{R3, R5}). 
We sign each claim with a unique private deployment key of the Advocate instance. 
This way all evidence is directly authenticated and after being anchored in the blockchain also immutable and publicly available. Hence, supporting later auditing of collected evidence.

Since the anchoring and storing of all raw evidence publicly is costly and possibly breaching confidentiality Advocate also offeres the \textit{Claim-Aggregation} controller (\textbf{R4}).
It is able to perform aggregation of evidence trails, either to prove specific evidence relationships publicly about all claims in a given period or to publish a claim that keeps the underlying claims data confidential.
This aggregation controller is currently able to run arbitrary, signed, Rego-Policies\footnote{https://www.openpolicyagent.org/docs/latest/policy-language/}.
The result of an aggregation is then stored and anchored in the same way as raw evidence (\textbf{R5}).
The aggregation controller is configurable and can easily be extended to use other methods for adding new properties to the claims, for example to support zero-knowledge proofs frameworks.

Advocate also follows a specific life cycle. First, during startup the identity of Advocate itself is claimed, which is always anchored.
This self-claim contains a unique name, the Advocate version, metadata about the environment (e.g., the location, configuration of the Kubernetes cluster) and a reference to the user that started Advocate.
All other evidence created thereafter is always referencing the self-claim, thus, linking the collected evidence always back to the specific instance generating them and also to the user that authorized the starting of the agent. 
Following this bootstrapping process, we start to continuously collect and aggregate evidence as it is configured.
In parallel, Advocate also provides an \textit{API} that allows any service in the cluster to generate and publish evidence claims directly. 
For this, a service provider must first obtain a publishing key from the Advocate operator, to ensure that responsible persons for the generated evidence are on record.
The API can then be used, to add additional context to an evidence trail that goes beyond the evidence collected through infrastructure and runtime services, e.g., specific businesses events. 
Lastly, Advocate can also enable the connection and collaboration with other Advocate instances to build evidence trails in multi-party organizations.\label{sec:architecture}
\section{Conclusion}\label{sec:concl}In this paper, we presented Advocate, a first approach to provide attested runtime evidence of cloud-native applications. 
We designed Advocate to enable a holistic integration into cloud-native runtime environments, while offering independent variability through an SII-like attestation of observability data (evidence).
Through this, aggreagated evidence can be combined with other evidence sources during the development phase to provide service consumers a comprehensive and traceable way to verify service delivery\cite{trustOps}.



\bibliographystyle{src/IEEEtran}
\bibliography{refs}

\end{document}